\documentclass[preprintnumbers]{revtex4}
\usepackage{graphicx}
\usepackage{dcolumn}
\usepackage{bm}
\def\sla#1{\slash \hspace{-2.5mm} #1}

\newcommand{\Del}{$\Delta$}
\long\def\Omit#1{}

\def\beq{\begin{equation}}
\def\eeq{\end{equation} }
\def\bea{\begin{eqnarray}}
\def\eea{\end{eqnarray}}
\def\eqref#1{Eq.~(\ref{eq:#1})}
\def\eqlab#1{\label{eq:#1}}
\def\figref#1{Fig.~(\ref{fig:#1})}
\def\figlab#1{\label{fig:#1}}

\def\tablab#1{\label{tab:#1}}
\def\secref#1{Section~\ref{sec:#1}}
\def\seclab#1{\label{sec:#1}}
\def\VYP#1#2#3{{\bf #1}, #3 (#2)}  

\def\NPA#1#2#3{Nucl.~Phys.~A~\VYP{#1}{#2}{#3}}
\def\NPB#1#2#3{Nucl.~Phys.~B~\VYP{#1}{#2}{#3}}

\def\PLB#1#2#3{Phys.~Lett.~B~\VYP{#1}{#2}{#3}}

\def\PRC#1#2#3{Phys.~Rev.~C~\VYP{#1}{#2}{#3}}
\def\PRD#1#2#3{Phys.~Rev.~D~\VYP{#1}{#2}{#3}}
\def\PRL#1#2#3{Phys.~Rev.~Lett.~\VYP{#1}{#2}{#3}}

\def\bge{\begin{equation}}
\def\ene{\end{equation}}
\def\bgea{\begin{eqnarray}}
\def\enea{\end{eqnarray}}

\begin{document}
\preprint{KVI/emi2001}
\title{Compton Scattering on the Proton.}

\author{O.\ Scholten} 
\affiliation{Kernfysisch Versneller Instituut, 
University of Groningen, 9747 AA Groningen, The~Netherlands}
\author{S.\ Kondratyuk} 
\affiliation{TRIUMF, 4004 Wesbrook Mall, Vancouver, 
British Columbia, Canada V6T 2A3}


\begin{abstract}
A microscopic coupled-channels model for Compton and pion 
scattering off the nucleon is introduced which is applicable at the 
lowest energies (polarizabilities) as well as at GeV energies. To 
introduce the model first the  conventional K-matrix approach is 
discussed to extend this in a following chapter to the ``Dressed K-Matrix" model. 
The latter approach restores causality, or analyticity, of the amplitude 
to a large extent. In particular, crossing
symmetry, gauge invariance and unitarity are satisfied. The extent of
violation of analyticity (causality) is used as an expansion parameter.
\end{abstract}
\maketitle

\section{Introduction}

In a K-matrix approach an infinite series of rescattering loops is taken 
into account with the approximation of incorporating only the pole 
contributions of the loop diagrams. In this approximation lies both the 
strong and the weak sides of this approach. By including only the pole 
contributions, which correspond to rescattering via physical states, 
unitarity in obeyed and the infinite series can be expressed as a 
geometric series and summed as given in \eqref{T-K}. Such an approach 
can be formulated in a co-variant approach, where electromagnetic-gauge invariance is 
obeyed through the addition of appropriate contact terms. It can also be 
shown that crossing symmetry is obeyed. In the structure of the kernel a 
large number of resonances can be accounted for. The weak point of 
including only the pole contributions of the loop corrections (i.e.\ 
only the imaginary part of the loop integrals) is that the resulting 
amplitude violates analyticity constraints and thus causality.

In \secref{Dressed} an extension of the K-matrix model is discussed 
where, without violating the other
symmetries, an additional constraint, that of analyticity (or
causality), is incorporated approximately. In this approach, called the ``Dressed
K-Matrix Model", dressed
self-energies and form factors are included in the K-matrix. These
functions are calculated self-consistently in an iteration procedure
where dispersion relations are used at each recursion step to relate
real and imaginary parts.

\section{The K-Matrix approach}
\seclab{K-Mat}

In K-matrix models the T-matrix is written in the form,
\beq
{\mathcal T}= (1- K i\delta)^{-1}\,K \; ,
\eqlab{T-K}
\end{equation}
where $\delta$ indicates that the intermediate particles have to be
taken on the mass shell and all physics is put in the kernel, the
K-matrix. This amounts to re-sum an infinite series of pole contributions 
ofloop corrections. 
It is straightforward to check that $S=1+2i{\mathcal{T}}$ is
unitary provided that the kernel $K$ is Hermitian. Since \eqref{T-K}
involves integrals only over on-shell intermediate particles, it reduces
to a set of algebraic equations when one is working in a partial wave
basis. When both the $\pi - N$ and $\gamma - N$ channels are open, the
coupled-channel K-matrix becomes a $2 \times 2$ matrix in the channel
space, i.e.\
\beq
K=\left[
\begin{array}{cc}
K_{\gamma \gamma} & K_{\gamma \pi} \\
K_{\pi \gamma} & K_{\pi \pi}
\end{array}
\right].
\eqlab{k-matr}
\end{equation}
It should be noted that due to the coupled channels nature of this
approach the widths of resonances are generated dynamically.

\begin{figure}[htb]
  \begin{center}
\includegraphics[height=0.10\textheight]{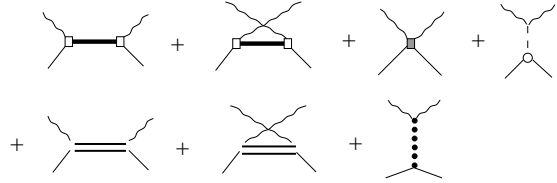}
\caption[f5]{The sum of diagrams included in the K matrix for
Compton scattering. 
A full spectrum of baryon resonances has been included.
 \figlab{K-gg}}
    \end{center}
\end{figure}

In traditional K-matrix models the kernel, the K-matrix, is built from
tree-level diagrams\cite{Gou94,Sch96,Feu98,Kor98}. In the present
investigation the type of diagrams included in $K_{\gamma \gamma}$ are
similar to that of ref.~\cite{Kor98} except that the \Del\ is treated
now as a genuine spin-3/2 resonance \cite{Pas01} in order to be
compatible with the later treatment of the in-medium \Del\ resonance.
This K-matrix is indicated in \figref{K-gg}. Most of the (non-strange)
resonances below 1.7 GeV have been included. The different coupling
constants were fitted to reproduce pion scattering, pion photoproduction
and Compton scattering on the nucleon. A comparable fit to the data as
in ref.~\cite{Kor98} could be obtained. In~\figref{n-comp} the results
for Compton scattering are compared to data.

\begin{figure}[hbt]
\begin{center}
\includegraphics[height=0.40\textheight]{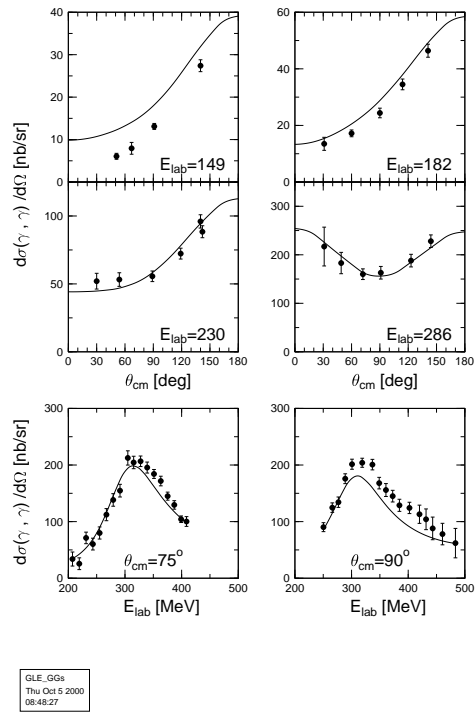}
\caption[dns]{The calculated cross section for Compton scattering off
the proton as a function of angle at fixed photon
energy, and as a function of photon energy at fixed angle. Data are taken
from ref.~\cite{Hal93}.
}
\figlab{n-comp}
\end{center}
\end{figure}

\section{Basic symmetries} 

A realistic scattering amplitude for a particular process should obey
certain symmetry relations, such as Unitarity, Covariance, Gauge
invariance, Crossing and Causality. In the following each of these
symmetries will be shortly addressed, stating its physical significance.
It is also indicated which of these is obeyed by the K-matrix approach
discussed in the previous section. The comparative success of the
K-matrix formalism can be regarded as due to the large number of
symmetries which are being obeyed. A violation of anyone of these
symmetries will directly imply some problems in applications.
Improvements are thus important.

\subsection{Unitarity} 

The unitarity condition for the scattering matrix $S$ reads $S^\dagger
S=1$. Usually one works with the $T$-matrix operator which can be
defined as $ S = 1 + 2\,i\,T$, and the unitarity condition is rewritten
as $2 i ( T T^\dagger )_{fi} = T_{fi} - T_{if}^*$. If the $T$-matrix is
symmetric (which is related to time-reversal symmetry),
the last formula becomes $ Im \ T_{fi}= \sum_n T_{fn} T_{in}^*$, where
the sum runs over physical intermediate states. The latter relation is
the generalization of the well-known optical theorem for the scattering
amplitude. Unitarity can only be obeyed in a coupled channel
formulation; the imaginary part of the amplitude ``knows" about the flux
that is lost in other channels.

In the K-matrix formalism the $T$-matrix is expressed as $T={K \over
1-i\,K}$ which implies that $S={1+i\,K \over 1-i\,K}$ is clearly unitary
provided the the kernel $K$ is Hermitian. This kernel is a matrix, where
the different rows and columns correspond to different physical outgoing
(incoming) channels. The coupled channels nature is thus inherent in
such an approach. As explained earlier the kernel is usually written as
the sum of all possible tree-level diagrams. In a partial-wave basis $K$
is a matrix of relatively low dimensionality and the inverse, implied in
the calculation of the $T$ matrix, can readily be calculated.

\subsection{Covariance} 

The scattering amplitude is said to be covariant if it transforms
properly under Lorentz transformations. As a consequence the description
of the reaction observables is independent of the particular reference
frame chosen for the calculations. It naturally implies that
relativistic kinematics is used.

Since the appropriate four-vector notation and $\gamma$-matrix algebra
are used throughout our calculation, the condition of Lorentz covariance is fulfilled.

\subsection{Gauge invariance} 

Gauge invariance means that there is certain freedom in the choice of
the electromagnetic field, not affecting the observables. Its
implication is current conservation, $\nabla \vec{J}={\partial \rho
\over \partial t}$, or in four-vector notation, $\partial_\mu J^\mu=0$.
Using the well known correspondence between momenta and derivatives,
current conservation can be re expressed as $k_\mu J^\mu=0$. If the
electromagnetic current obeys this relation it can easily be shown that
observables, such as a photo-production cross section, are independent
of the particular gauge used for constructing the photon polarization
vectors.

One of the sources for violation of gauge invariance is the form factors
used in the vertices. A form factor implies that at a certain (short
range) scale a particle appears 'fuzzy'. At distances smaller than this
scale deviations from a point-like structure are important; however in
the formulation the dynamics at this short scale is not sufficiently
accurate. For one thing, the flow of charge at this scale is not
properly accounted for, implying violation of charge conservation. To
correct for this, so-called contact terms are usually included in the
$K$-matrix kernel. In the present model these contact terms are
constructed using the minimal substitution rules. The corresponding T-matrix,
as well as the observables, are independent of the photon gauge.

\subsection{Crossing Symmetry} 

Physical consequences of the crossing symmetry are more difficult to
explain. It basically means that in a proper field-theoretical framework
the scattering amplitudes of processes in the so-called crossed channels
can be obtained from each other by appropriate replacements of
kinematics. This assumes that the amplitude can be analytically
continued from the physical region of one channel to the physical
regions of other channels. An example of the crossed channels is $\gamma
N \to \pi N$, $\pi N \to \gamma N$ and $N \bar{N} \to \gamma \pi$. 
\Omit{In
particular the property of detailed balance follows from the relation
between the first two reactions in this chain (the so-called s-u channel
crossing). 
\bf ! Alex: The last sentence is not completely clear.
Detailed balance usually follows from the time inversion. Are you sure
the s-u crossing leads to the same result? Actually, the 2nd reaction in
the chain I'd write as $\gamma \bar{N} \to \pi \bar{N}$ though what you
write seems also correct. }

Crossing symmetry puts a direct constraint on the amplitude for the case
that direct and crossed channels are identical, as for example for the
processes $\pi N \to \pi N$ or $\gamma N \to \gamma N$. In these
reactions crossing symmetry leads to important symmetry properties of
the amplitudes under interchange of $s$ and $u$ variables. Due to the
fact that in the $K$-matrix formalism the rescattering diagrams which
are taken into account have only on-shell intermediate particles, it can
be shown that the s-u crossing symmetry is obeyed provided that the
kernel itself is crossing symmetric. Since the latter is the case,
crossing symmetry is obeyed.

\subsection{Analyticity} 

Analyticity of the scattering matrix is not really a symmetry. Rather, 
it requires that the amplitude be an analytic function of the energy 
variable and in particular that it obeys dispersion relations. The physics 
of analyticity is closely related to causality of 
the amplitude as is illustrated in the following example. 

Assume that a signal is emitted by an antenna at time $t=0$. At all
subsequent times the signal is given by a function $F(t)$ while
causality requires that at earlier times there was no signal,
$F(t<0)=0$. This signal can be Fourier-transformed,
$f(\omega)=\int_{0}^{+\infty}\! dt\, e^{i \omega t} \, F(t)$ to
explicitly show its energy or frequency dependence. Note that the
integration region from $t=-\infty$ to $t=0$ gives zero contribution due
to the causality requirement. This transformation can also be considered
for complex values of $ \omega$. Since the integration interval runs
only over positive values for $t$ the Fourier integral exists and is a
well behaving function for all complex values of energy $\omega$ for
which $Im(\omega)>0$ i.e.\ it is an analytic function. For such a
function contour integrals in the complex $\omega$ plane can be
performed and the function obeys the Cauchy theorem which is in this
context is usually formulated as a dispersion relation,
\[
\mbox{Re} f(\omega)=\frac{\mathcal{P}}{\pi}\int_{-\infty}^{+\infty}
\! d \omega' \frac{\mbox{Im} f(\omega')}{\omega'-\omega}
\]
showing that for an analytic function the real and imaginary parts are
closely related. For example, if the imaginary part of an analytic
function is given by the curve on the left-hand side of \figref{cauchy}
the real part of this function is given by the right-hand side.

\begin{figure}[htb] 
  \begin{center} 
\begin{minipage}[t]{45mm} 
\includegraphics[height=4cm,bb=6 170 530 650,angle=0] 
{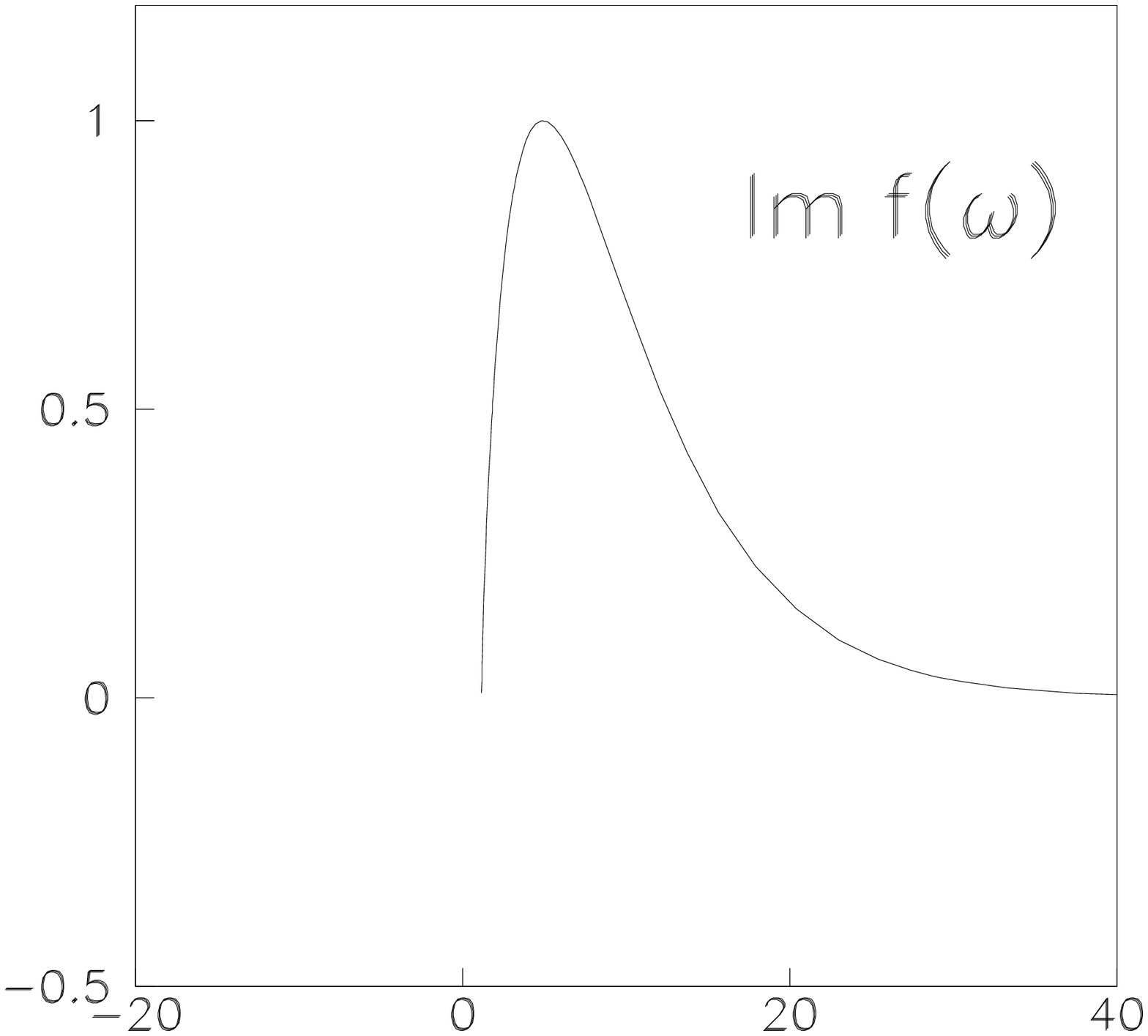}
\end{minipage} \ \ \ \ \ 
\begin{minipage}[t]{45mm} 
\includegraphics[height=4cm,bb=6 170 530 650,angle=0] 
{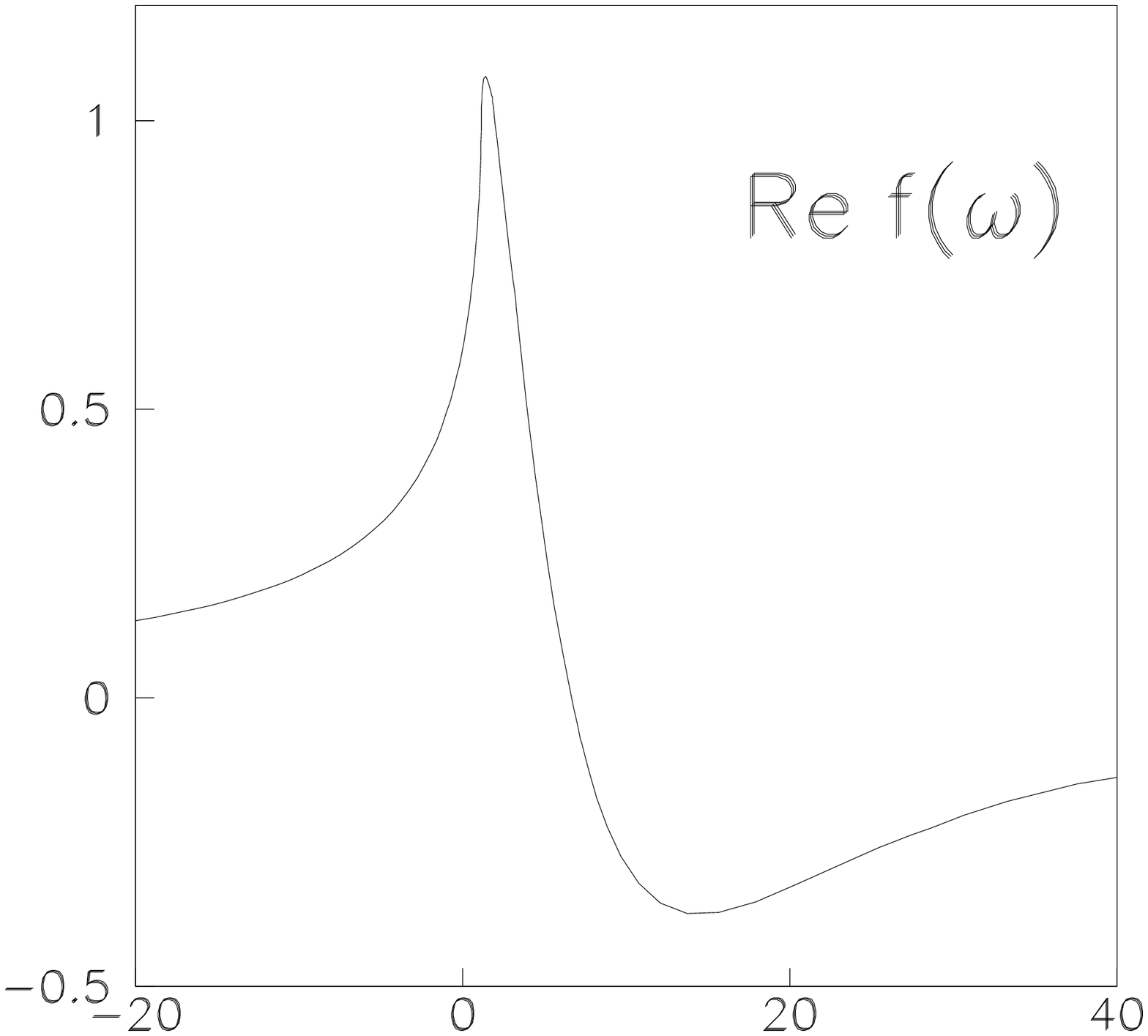} 
\end{minipage} 
    \caption[f1]{An example of the real and imaginary parts of an 
analytic function which are related through a dispersion relation. 
\figlab{cauchy}} 
    \end{center} 
\end{figure} 

In the traditionally used K-matrix approaches the analyticity constraint 
is badly violated. The origin of this is explained in 
the following. 
 
In a field-theoretical calculation of a scattering amplitude one
includes rescattering contributions of intermediate particles which are
expressed as loop integrals. In \figref{sigma-loop} a typical loop
contribution to the self energy is shown. Ignoring terms in the
numerator which are irrelevant for the analyticity properties, the
corresponding integral can be expressed as
\beq
J(p^2)= \int \!d^4 k \,\frac{1}{[k^2-\mu^2+i \epsilon]\,
[(p-k)^2-m^2+i \epsilon]} = \mbox{Re} J(p^2) + i \mbox{Im} J(p^2)
\eqlab{sigma-loop}\end{equation}
where the right hand side in this equation and in \figref{sigma-loop}
expresses the fact this integral has a real and an imaginary part, each
of which corresponds to some particular physics. The imaginary part of
the integral arises from the integration region where the denominators
vanish, corresponding to four-momenta $k$ where the intermediate
particles in the loop are on the mass shell -or equivalently- are
physical particles with $k^2=\mu^2$ and $(p-k)^2=m^2$.
Conventionally this is indicated by placing a
slash through the loop (see \figref{sigma-loop}) to indicate that the
loop can be cut at this place since it corresponds to a physical state.
The other parts of the integration region contribute to the real part of
the integral. In the latter case the particles in the loop are off the
mass shell.

\begin{figure}[htb] 
  \begin{center} 
\includegraphics[height=2cm,bb=41 360 550 480,angle=0] 
{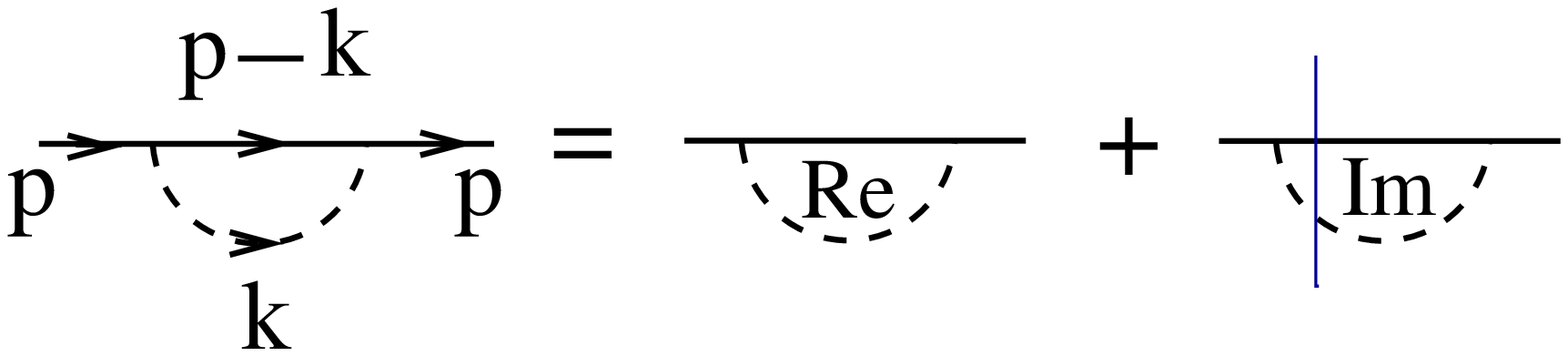} 
    \caption[f1]{Loop integral contributing to the self energy. 
\figlab{sigma-loop}} 
    \end{center} 
\end{figure} 

It can be shown that the K-matrix formulation for the T-matrix
corresponds to including only the imaginary (or cut-loop) contributions
of a certain class of loop diagrams. This guarantees (as was shown
before) that unitarity is obeyed. Analyticity of the scattering
amplitude is however violated due to ignoring the real contributions of
these loop integrals. As a consequence causality will be violated!

To (partially) recover analyticity of the scattering amplitude the so-called 
``Dressed K-matrix approach''\cite{Kon01b} has been developed. 
It is described in the following section.

\section{The Dressed K-matrix Model} \seclab{Dressed}

As discussed in the previous section, the coupled channels K-matrix
approach is quite successful in reproducing Compton scattering. However
it fails in predicting nucleon polarizabilities. The reason is that, in
spite of the many symmetry properties that are satisfied, analyticity or
causality of the amplitude is badly violated. In the ``Dressed K-matrix"
approach the constraint of analyticity is incorporated in an approximate
manner without spoiling the other symmetries. In fact analyticity is
used as a kind of expansion parameter where presently only the leading
contributions are included. The ingredients of the Dressed K-Matrix
Model were described in Refs.~\cite{Kon99,Kon00,Kon01a} and the main
results were presented in Ref.~\cite{Kon01b}. The essence of this
approach lies in the use of {\em dressed} vertices and propagators in
the kernel $K$.

The objective of dressing the vertices and propagators is solely to
improve on the analytic properties of the amplitude. The imaginary parts
of the amplitude are generated through the K-matrix formalism (as
imposed by unitarity) and correspond to cut loop corrections where the
intermediate particles are taken on their mass shell. The real parts
have to follow from applying dispersion relations to the imaginary
parts. We incorporate these real parts as real vertex and self-energy
functions. Investigating this in detail (for a more extensive discussion
we refer to ~\cite{Kon99}) shows that the dressing can be formulated in
terms of coupled equations, schematically shown in \figref{piNN}, which
generate multiple overlapping loop corrections. The coupled nature of
the equations is necessary to obey simultaneously unitarity and
analyticity at the level of vertices and propagators.

\begin{figure}[htb]
\begin{center}
\includegraphics[height=0.40\textheight]{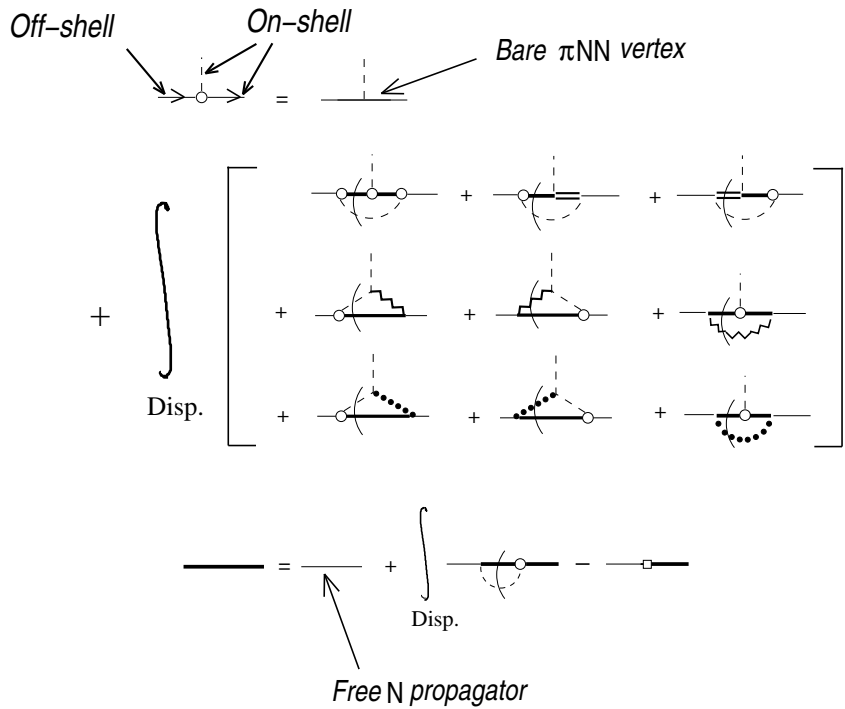}
\caption[f1]{Graphical representation of the equation for the dressed
irreducible $\pi N N$ vertex, denoted by an open circle, and the dressed
nucleon propagator, denoted by a solid line. The dashed lines denote
pions, the double lines denote $\Delta$s and the zigzag and dotted lines
are $\rho$ and $\sigma$ mesons, respectively. The resonance propagators
are dressed. The last term in the second equation denotes the
counter-term contribution to the nucleon propagator, necessary for the
renormalization. \figlab{piNN}}
\end{center}
\end{figure}

The equations presented in \figref{piNN} are solved by iteration where
every iteration step proceeds as follows. The imaginary -- or pole --
contributions of the loop integrals for both the propagators and the
vertices are obtained by applying cutting rules. Since the outgoing
nucleon and the pion are on-shell, the only kinematically allowed cuts
are those shown in \figref{piNN}. The principal-value part of the vertex
(i.e. the real parts of the form factors) and self-energy functions are
calculated at every iteration step by applying dispersion relations to
the imaginary parts just calculated, where only the physical
one-pion--one-nucleon cut on the real axis in the complex $p^2$-plane is
considered. These real functions are used to calculate the pole
contribution for the next iteration step. This procedure is repeated to
obtain a converged solution. We consider irreducible vertices, which
means that the external propagators are not included in the dressing of
the vertices.

Bare $\pi NN$ form factors have been introduced in the dressing
procedure to regularize the dispersion integrals. The bare form factor
reflects physics at energy scales beyond those of the included mesons
and which has been left out of the dressing procedure. One thus expects
a large width for this factor, as is indeed the case.

The dressed nucleon propagator is renormalized (through a wave function 
renormalization factor $Z$ and a bare mass $m_0$) to have a pole with a unit
residue at the physical mass. The nucleon self-energy is expressed in 
terms of self-energy functions $A(p^2)$ and $B(p^2)$ as $
\Sigma_N(p)= A_N(p^2)\;\sla{p} + B_N(p^2)\;m$.

The procedure of obtaining the $\gamma N N$ vertex~\cite{Kon00} is in
principle the same as for the $\pi N N$ vertex. Contact $\gamma \pi N N$
and $\gamma \gamma N N$ vertices, necessary for gauge invariance of the
model, are constructed by minimal substitution in the dressed $\pi N N$
vertex and nucleon propagator, as was explained in \cite{Kon00}.

The present procedure restores analyticity at the level of one-particle
reducible diagrams in the T-matrix. In general, violation due to two-
and more-particle reducible diagrams can be regarded as higher order
corrections. An important exception to this general rule is formed by,
for example, diagrams where both photons couple to the same intermediate
pion in a loop (so-called ``handbag" diagrams). This term is exceptional
since at the pion threshold the S-wave contribution is large, due to the
non-zero value of the $E_{0+}^{1/2}$ multipole in pion-photoproduction,
leading to a sharp near-threshold energy dependence of the related
$f_{EE}^{1-}$ Compton amplitude\cite{Ber93}. In the K-matrix formalism,
the imaginary (pole) contribution of this type of diagrams is taken into
account. Not including the real part of such a large contribution would
entail a significant violation of analyticity. To correct for this, the
$\gamma \gamma N N$ vertex also contains the (purely transverse) ``cusp"
contact term whose construction is described in Section 4 of
Ref.~\cite{Kon00}. Since, due to chiral symmetry, the S-wave pion
scattering amplitude vanishes at threshold, the mechanism that gives 
rise to the important ``cusp" term in compton scattering does not 
contribute to $\pi\pi N N$ or $\pi \gamma N N$ contact terms. The  
analogons to the ``cusp" $\gamma \gamma N N$ term
will thus be negligible and have therefore not been considered.

\subsection{Results }  \seclab{results}

Results for pion-nucleon scattering and pion-photoproduction obtained in
the dressed K-matrix model and in the traditional K-matrix approach are
of comparable quality. One should, however, expect the two approaches to
have significant differences for Compton scattering since for this case
constraints imposed by analyticity will be most
important~\cite{Pfe74,Ber93}.

The effect of the dressing on the $f_{EE}^{1-}$ amplitude can be seen in
\figref{fee}, where also the results of dispersion analyses are quoted
for comparison. Note that the imaginary parts of $f_{EE}^{1-}$ from
calculations B (Bare, corresponding to the usual K-matrix approach) and 
D (Dressed, the full Dressed K-matrix results) are rather similar in the vicinity of threshold.

\begin{figure}[htb]
\begin{center}
\includegraphics[height=0.30\textheight]{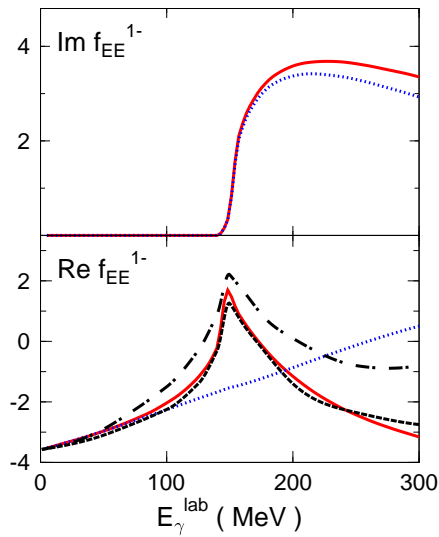}
\caption[f14]{The
$f_{EE}^{1-}$ partial amplitude of Compton scattering on the proton in
units $10^{- 4}/m_{\pi}$. Solid line: dressed K-matrix, D; dotted line:
bare K-matrix, B. Also shown are the results of the dispersion analyses of
Ref.~\cite{Pfe74} (dash-dots) and Ref.~\cite{Ber93} (dashed).
\figlab{fee}}
\end{center}
\end{figure}

The polarizabilities characterize response of the nucleon to an externally
applied electromagnetic field \cite{Hem98,Hol00}. They can be defined as
coefficients in a low-energy expansion of the cross section or partial
amplitudes of Compton scattering. Since gauge invariance, unitarity,
crossing and CPT symmetries are fulfilled in both models the Thompson limit
at vanishing photon energy is reproduced. Our results for the electric,
magnetic and spin polarizabilities of the proton are given in
Table~\ref{tab:polp}, where they are compared with the results given in
refs.~\cite{Hem98,Gel00} and with the values extracted from recent
experiments. The contribution from the the t-channel $\pi^0$-exchange
diagram has been subtracted. The effect of the dressing on the
polarizabilities can be seen by comparing the values given in columns D 
(dressed)
and B (bare). In particular, the dressing tends to decrease $\alpha$ while
increasing $\beta$. Among the spin polarizabilities, $\gamma_{E1}$ is
affected much more than the other $\gamma$'s. The effect of the additional
``cusp'' $\gamma \gamma N N$ contact term\cite{Kon00}, strongly
influences the electric polarizabilities rather than the magnetic ones.
This is because the ``cusp" contact term affects primarily the
electric partial amplitude $f_{EE}^{1-}$ (corresponding to the the total
angular momentum and parity of the intermediate state $J^\pi=1/2^-$)
rather than the magnetic amplitude $f_{MM}^{1-}$ ($J^\pi=1/2^+$).

\begin{table}[htb]
\caption[T4]{Polarizabilities of the proton. The units are
$10^{-4}fm^3$ for $\alpha$ and $\beta$ and $10^{-4}fm^4$ for the
$\gamma$'s (the anomalous $\pi^0$ contribution is subtracted).
The first two columns contain the polarizabilities obtained from the
present calculation; D (full, dressed) and B (bare K-matrix).
The two columns named $\chi PT$ contain the
polarizabilities calculated in the chiral perturbation 
theory~\cite{Hem98,Gel00}.
Results of recent dispersion analyses are given in the last column
(Ref.~\cite{Lvo97} for $\alpha$ and $\beta$ and Ref.~\cite{Dre00} for the
$\gamma$'s). } 
\begin{center}
\begin{tabular}{|c|cc|cc|c|}
\hline
& & & 
{ $\chi_{PT}$}   & & {DA} \\
               &{ D}&{ B}& Gel00 & Hem98  & \\
\hline 
$\alpha$    &{12.1}&{15.5} & 10.5 &16.4 &11.9  \\ 
$\beta$     &{ 2.4}&{ 1.7} & 3.5  & 9.1 & 1.9  \\
$\gamma_{E1}$ &{-5.0}&{-1.7} & -1.9 &-5.4 &-4.3 \\
$\gamma_{M1}$ &{ 3.4}&{ 3.8} & 0.4  & 1.4 & 2.9 \\
$\gamma_{E2}$ &{ 1.1}&{ 1.0} & 1.9  & 1.0 & 2.2 \\
$\gamma_{M2}$ &{-1.8}&{-2.3} & 0.7  & 1.0 & 0.0 \\
$\gamma_0$    &{ 2.4}&{-0.9} &-1.1  & 2.0 &-0.8 \\
$\gamma_\pi$  &{11.4}&{ 8.9} & 3.5  & 6.8 & 9.4 \\
\hline
\end{tabular}

\end{center}
\tablab{polp}
\end{table}

Of special interest is to check whether polarizabilities as extracted form the 
low energy behavior of the amplitude are in agreement with the values as 
extracted from energy weighted sumrules. The derivation of sumrules is 
based on the fact that the amplitudes obey certain symmetries where 
analyticity is of particular importance. This comparison is still in 
progess and results will be published in a forthcoming paper 
\cite{Kon01c}. Preliminary results indicate that the different sum rules 
are obeyed, with the exception of the sum rule for the spin 
polarizability $\gamma_0$. This may be due to an incomplete dressing of 
the $\Delta$-resonance in the present calculational scheme.

\section{Summary and Conclusions}

Results are presented for Compton scattering on a free proton as well as 
on a nucleus. For processes on the proton the usual K-matrix model as well as the 
recently developed dressed K-matrix model are discussed. In the latter 
approach the real self energies and vertex functions are obtained from the 
imaginary parts using dispersion relations imposing self-consistency 
conditions. It is indicated that such an 
approach is essential to understand features seen in the data, in 
particular at energies around and below the pion production threshold.

\section*{Acknowledgments}

O.~S.\ thanks the Stichting voor Fundamenteel Onderzoek der Materie (FOM) for
their financial support.
S.~K.\ thanks the National Sciences and Engeneering Research 
Council of Canada for their financial support.

\end{document}